# Fluctuations of Quantum Entanglement

E. B. Fel'dman and M. A. Yurishchev

*Institute of Problems of Chemical Physics, Russian Academy of Sciences,
Chernogolovka, Moscow region, 142432 Russia*
*e-mail: efeldman@icp.ac.ru, yur@itp.ac.ru*

It is emphasized that quantum entanglement determined in terms of the von Neumann entropy operator is a stochastic quantity and, therefore, can fluctuate. The rms fluctuations of the entanglement entropy of two-qubit systems in both pure and mixed states have been obtained. It has been found that entanglement fluctuations in the maximally entangled states are absent. Regions where the entanglement fluctuations are larger than the entanglement itself (strong fluctuation regions) have been revealed. It has been found that the magnitude of the relative entanglement fluctuations is divergent at the points of the transition of systems from an entangled state to a separable state. It has been shown that entanglement fluctuations vanish in the separable states.

Quantum entanglement ($E$) has been actively investigated both theoretically and experimentally in recent time [1–4]. The entropy of a state with a reduced (partial) density matrix appeared to be one of the most productive and deep measures of entanglement. Such an entropy was considered in 1986 in the problem of black holes [5]. The entropy of the reduced state is widely used in various fields including quantum field theory, solid state physics, and, certainly, quantum information (see, e.g., [6] and references cited therein).

The physical (information) meaning of entanglement determined in terms of entropy is treated as the relative number of maximally entangled pairs $m$ that can be extracted from a large number $n$ of copies of the initial systems using a cleaning protocol including only local operations and classical communication: $E \longrightarrow m/n$, $n \longrightarrow \infty$ [7, 8]. However, this value is only the mean (expectation) value of entanglement.

Relating quantum entanglement with entropy, which, as is known, can fluctuate (see [9, Sect. 112]), one should take into account the possibility of entanglement fluctuations. For a large set of statistically independent subsystems, the volume-averaged relative fluctuations of the physical characteristics are negligibly small, because they decrease as $1/\sqrt{n}$ (see [9, Sect. 2]). However, as will be shown, entanglement fluctuations for single composed systems can be significant and even reach infinitely large values. This work is devoted to the calculation and discussion of the features of the behavior of such fluctuations of entanglement entropy.

**Pure states.** According to [8], the measure of the entanglement of the system that is in a pure state $|\psi\rangle$ and consists of two subsystems $A$ and $B$ can be the von Neumann entropy of any of the subsystems

$$E = \bar{S}(\rho_A) = \bar{S}(\rho_B). \tag{1}$$

Here, $\rho_A = \text{Tr}_B |\psi\rangle\langle\psi|$ and $\rho_B = \text{Tr}_A |\psi\rangle\langle\psi|$ are the reduced density matrices, $\bar{S} = \text{Tr}\rho S$, where $S = -\log_2\rho$ is the entropy operator and $\rho \in \{\rho_A, \rho_B\}$.

Operators correspond to physical quantities in quantum mechanics. We identify the quantum entanglement operator $\hat{E}$ with the entropy operator $\hat{E} = S$. The operator $\hat{E}$ is equivalent to the "entanglement Hamiltonian" [10, 11].

As mentioned above, the entropy of the subsystems can have fluctuations; hence, they should not be excluded a priori for entanglement. According to the general definition of rms fluctuations of a random variable [9, Sect. 2], their magnitude is

$$\Delta E = [\bar{S}^2 - (\bar{S})^2]^{1/2}, \tag{2}$$

where the moments $\bar{S}$ and $\bar{S}^2$ are taken for one of the subsystems, $A$ or $B$.

Let us consider the behavior of quantum entanglement fluctuations in a two-qubit model. For this, the most general wavefunction represented in terms of the standard basis has the form

$$|\psi\rangle = a|00\rangle + b|01\rangle + c|10\rangle + d|11\rangle, \tag{3}$$

where $|a|^2 + |b|^2 + |c|^2 + |d|^2 = 1$. Therefore,

$$\rho = |\psi\rangle\langle\psi| = \begin{pmatrix} |a|^2 & ab^* & ac^* & ad^* \\ a^*b & |b|^2 & bc^* & bd^* \\ a^*c & b^*c & |c|^2 & cd^* \\ a^*d & b^*d & c^*d & |d|^2 \end{pmatrix}. \quad (4)$$

One eigenvalue of this matrix is $|a|^2 + |b|^2 + |c|^2 + |d|^2 (= 1)$ and the vector $|\psi\rangle$ corresponds to this eigenvalue. The three other eigenvalues are zero. According to Eq. (4), the reduced density matrices are represented as

$$\rho_A = \begin{pmatrix} |a|^2 + |b|^2 & ac^* + bd^* \\ a^*c + b^*d & |c|^2 + |d|^2 \end{pmatrix},$$

$$\rho_B = \begin{pmatrix} |a|^2 + |c|^2 & ab^* + cd^* \\ a^*b + c^*d & |b|^2 + |d|^2 \end{pmatrix}. \quad (5)$$

Taking into account the normalization $\langle\psi|\psi\rangle = 1$, the eigenvalues of each of these matrices are

$$\lambda_{1,2} = \frac{1}{2}(1 \pm \sqrt{1-C^2}), \quad (6)$$

where

$$C = 2|ad - bc|. \quad (7)$$

For the $k$th moment of the entropy operator of a subsystem (qubit $A$ or $B$), simple calculations yield

$$\bar{S}^k = (-1)^k \left[ \frac{1+\sqrt{1-C^2}}{2} \log_2^k\left(\frac{1+\sqrt{1-C^2}}{2}\right) + \frac{1-\sqrt{1-C^2}}{2} \log_2^k\left(\frac{1-\sqrt{1-C^2}}{2}\right) \right]. \quad (8)$$

Thus, all of the characteristics of the entanglement of the two-qubit system are concentrated only in one quantity, concurrence $C$.

From Eq. (8), entanglement (the first moment of entropy) is expressed as

$$E = H((1+\sqrt{1-C^2})/2), \quad (9)$$

where $H(x) = -x\log_2 x - (1-x)\log_2(1-x)$ is the Shannon function. Near the boundary points,

$$E(C) = \begin{cases} -\frac{1}{2}C^2 \log_2[C/(2\sqrt{e})], & C \to 0 \\ 1 - (1-C)/\ln 2, & C \to 1. \end{cases} \quad (10)$$

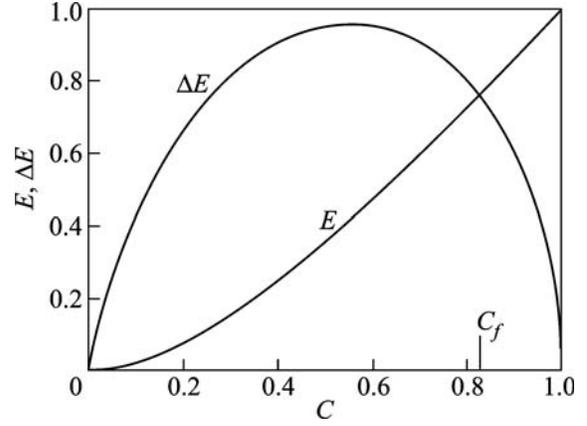

**Fig. 1.** Entanglement $E$ and its fluctuations $\Delta E$ in the two-qubit system in a pure state. The value $C_f = 0.82724...$ is marked in the abscissa axis.

Using Eqs. (2) and (8), the fluctuations of the quantum entanglement of the two-qubit system are represented in the form

$$\Delta E = C \log_2\left[\frac{1}{C}(1+\sqrt{1-C^2})\right]. \quad (11)$$

According to this expression,

$$\Delta E(C) = \begin{cases} -C\log_2(C/2), & C \to 0 \\ (\sqrt{2}/\ln 2)(1-C)^{1/2}, & C \to 1. \end{cases} \quad (12)$$

Relations (9) and (11) in the parametric form provide the $\Delta E(E)$ dependence, which can be used for indirect measurements of $\Delta E$ in terms of $E$.

Figure 1 shows the quantum entanglement and its fluctuations as functions of $C$ in the system under consideration. Absolute fluctuations at the point of zero entanglement ($C = 0$) are absent: $\Delta E = 0$. Near this point, function (11) contains a logarithmic singularity (see Eq. (12)). When $C = 1$ (maximally entangled state), entanglement fluctuations are also absent. However, these fluctuations increase as concurrence deviates from unity (see Fig. 1). The fluctuations become equal to entanglement ($\Delta E = E$) at $C = C_f$, where $C_f$ is the root of the transcendent equation

$$(C_f + \sqrt{1-C_f^2})\ln\left[\frac{1}{C_f}(1+\sqrt{1-C_f^2})\right] = \ln\frac{2}{C_f}. \quad (13)$$

This root is $C_f \approx 0.82724$. At $0 < C < C_f$, the system is in the region of strong fluctuations where $\Delta E > E$.

As $C$ decreases from unity to zero, the magnitude of the relative quantum-entanglement fluctuations, $\delta E = \Delta E/E$, increases monotonically from zero to infinity (see Fig. 2). Near the boundaries of the $C$ variation

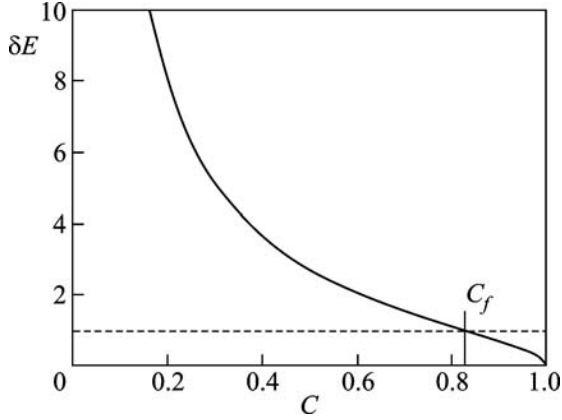

**Fig. 2.** (Solid line) Relative fluctuation versus $C$ for the two-qubit system in a pure state. The dashed straight line is the level $\delta E = 1$.

region, the relative entanglement fluctuations behave as

$$\delta E(C \longrightarrow 1) \simeq \frac{\sqrt{2}}{\ln 2}(1-C)^{1/2}, \quad (14)$$

$$\delta E(C \longrightarrow 0) \simeq \frac{2}{C}\left(1 - \frac{1}{2\ln(C/2)}\right)^{-1} \simeq 2C^{-1}. \quad (15)$$

Thus, the slope of the function $\delta E(C)$ at the boundary point $C = 1$ is infinite. In this region, entanglement fluctuations are very sensitive to variations of the system parameters. At the same time, the relative entanglement fluctuations tend to infinity near the point $C = 0$, at which the system transits from the entangled state to the separable one. This is well seen in Fig. 2.

**Mixed states.** Formation (creation) entanglement in the system in the state with the density matrix

$$\rho = \sum_i p_i |\psi_i\rangle\langle\psi_i| \quad (16)$$

(weights $p_i \geq 0$, $\sum_i p_i = 1$) is defined as [12]

$$E = \min_{\mathcal{E}} \sum_i p_i \bar{S}(\psi_i). \quad (17)$$

Here, $\bar{S}(\psi_i)$ is the entanglement of the pure state $|\psi_i\rangle$ (the method for its calculation for a two-component system is given above). A minimum in Eq. (17) should be found among all ensembles $\mathcal{E} = \{p_i, \psi_i\}$ with the conservation of the state $\rho$.

For a set of identical pairs of spin-1/2 systems, formation entanglement is determined by the minimum number of maximally entangled pairs that are necessary for the creation of a given state $\rho$ using local operations and classical communication [12].

Expression (17) for entanglement can be represented in the form

$$E = \langle \bar{S}(\psi_i)\rangle_{\mathcal{E}_o} = \sum_i p_i^o \bar{S}(\psi_i^o), \quad (18)$$

where $\mathcal{E}_o = \{p_i^o, \psi_i^o\}$ is the minimizing (optimal) ensemble.

As a measure of entanglement fluctuations in the mixed state, we take the standard deviation of entanglement in ensemble $\mathcal{E}_o$:

$$\Delta E = [\langle \bar{S}^2(\psi_i)\rangle_{\mathcal{E}_o} - \langle \bar{S}(\psi_i)\rangle_{\mathcal{E}_o}^2]^{1/2}. \quad (19)$$

Let us consider two-qubit systems. It was shown in [12–14] that all states $|\psi_i^o\rangle$ have the same concurrences (entanglements). Owing to this property of the optimal ensemble, all moments (8) are also the same. In particular, instead of Eq. (19), we have

$$\Delta E = [\bar{S}^2(\psi_i^o) - \bar{S}(\psi_i^o)^2]^{1/2}, \quad \forall i. \quad (20)$$

As a result, the magnitude of the entanglement fluctuations in the mixed state of the two-qubit system is as before given by Eq. (11). Thus, the calculation of the entanglement fluctuations in the two-qubit model reduces to the determination of concurrence.

The concurrence for an arbitrary two-qubit system can be calculated by the Hill–Wootters formula [13, 14] (see also [15, 16]):

$$C = \max\{0, \sqrt{\lambda_1} - \sqrt{\lambda_2} - \sqrt{\lambda_3} - \sqrt{\lambda_4}\}. \quad (21)$$

Here, $\lambda_i$ are the eigenvalues ($\lambda_1 \geq \lambda_2 \geq \lambda_3 \geq \lambda_4 \geq 0$) of the matrix

$$R = \rho(\sigma_y \otimes \sigma_y)\rho^*(\sigma_y \otimes \sigma_y), \quad (22)$$

where $\sigma_y$ is the Pauli matrix. Since the product of two noncommuting Hermitian matrices is a non-Hermitian matrix [17, Sect. 4], matrix $R$ is generally non-Hermitian. However, if $\det \rho \neq 0$, the similarity transformation reduces the $R$ matrix to the Hermitian form

$$R' = \rho^{-1/2}R\rho^{1/2} = \sqrt{\rho}(\sigma_y \otimes \sigma_y)\rho^*(\sigma_y \otimes \sigma_y)\sqrt{\rho}. \quad (23)$$

It is easy to verify that the corresponding $R$ matrix for the pure state with the density matrix given by Eq. (4) has only one nonzero eigenvalue $4|ad - bc|^2$ and Hill–Wootters formula (21) gives Eq. (7).

Another case corresponds to the state whose density matrix is an arbitrary mixture of the Bell states:

$$\rho = p_1|\Psi^+\rangle\langle\Psi^+| + p_2|\Psi^-\rangle\langle\Psi^-| \\ + p_3|\Phi^+\rangle\langle\Phi^+| + p_4|\Phi^-\rangle\langle\Phi^-|, \quad (24)$$

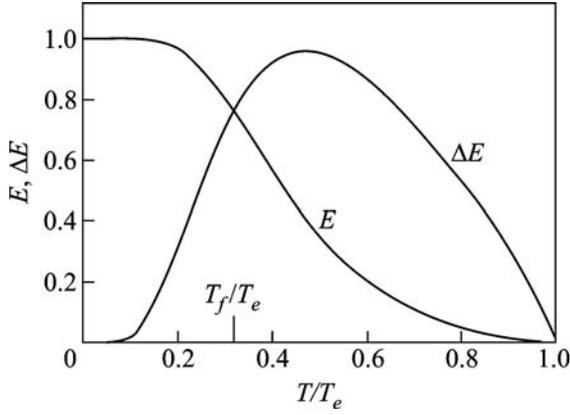

**Fig. 3.** Entanglement $E$ and its fluctuation $\Delta E$ in the Heisenberg dimer versus the dimensionless temperature $T/T_e$. The value $T_f/T_e = 0.31776...$ is marked in the abscissa axis.

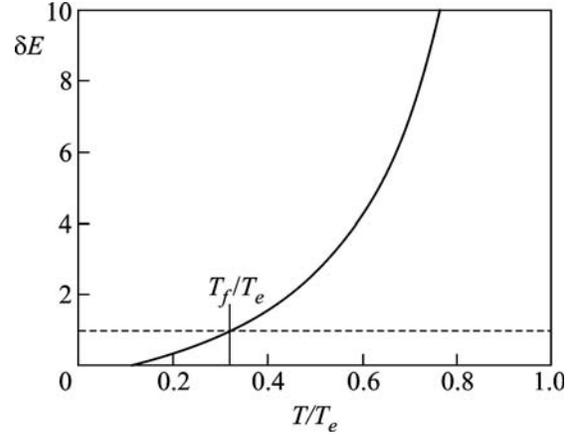

**Fig. 4.** (Solid line) $\delta E$ in the Heisenberg dimer versus the dimensionless temperature $T/T_e$. The dashed straight line is the level $\delta E = 1$.

where

$$|\Psi^\pm\rangle = \frac{1}{\sqrt{2}}(|01\rangle \pm |10\rangle), \quad |\Phi^\pm\rangle = \frac{1}{\sqrt{2}}(|00\rangle \pm |11\rangle) \quad (25)$$

are the Bell functions. Density matrix (24) has the structure

$$\rho = \frac{1}{2}\begin{pmatrix} p_3 + p_4 & & & p_3 - p_4 \\ & p_1 + p_2 & p_1 - p_2 & \\ & p_1 - p_2 & p_1 + p_2 & \\ p_3 - p_4 & & & p_3 + p_4 \end{pmatrix}. \quad (26)$$

For such a system, $R = \rho^2$ and the eigenvalues of the matrix $R$ are $p_i^2$. For this reason, the Hill–Wootters formula provides the concurrence

$$C = \begin{cases} 2p_{max} - 1, & p_{max} > 1/2 \\ 0, & p_{max} \leq 1/2, \end{cases} \quad (27)$$

where $p_{max} = \max\{p_1, p_2, p_3, p_4\}$. Relations (27) and (9) return us to the results obtained in [12].

Let us consider a Heisenberg dimer with the Hamiltonian

$$\mathcal{H} = -\frac{1}{2}J\sigma_1\sigma_2, \quad (28)$$

where $J$ is the coupling constant and $\sigma = (\sigma_x, \sigma_y, \sigma_z)$ is the vector of Pauli matrices (at site 1 or 2). The density matrix of the dimer that is in equilibrium with a thermostat at temperature $T$ has the form (see [18])

$$\rho(T) \equiv \frac{1}{Z}\exp(-\mathcal{H}/k_B T) = \frac{1}{Z}e^{-3K}|\Psi^-\rangle\langle\Psi^-| \\ + \frac{1}{Z}e^K(|\Psi^+\rangle\langle\Psi^+| + |\Phi^+\rangle\langle\Phi^+| + |\Phi^-\rangle\langle\Phi^-|), \quad (29)$$

where

$$Z = 3e^K + e^{-3K} \quad (30)$$

is the partition function, $K = J/2k_B T$, and $k_B$ is the Boltzmann constant.

Density matrix (29) is a partial case of Eq. (24) and corresponds to the Werner density matrix [12]. According to Eq. (27), the concurrence of such a system can be nonzero only in the antiferromagnetic case ($J < 0$). In this case, it is given by the expression [19, 20] (see also [16, 18])

$$C(T) = \begin{cases} -1 + 2/[1 + 3\exp(-2|J|/k_B T)], & T < T_e \\ 0, & T \geq T_e, \end{cases} \quad (31)$$

where

$$T_e = \frac{2}{\ln 3}|J|/k_B. \quad (32)$$

This expression and Eq. (11) provide the dependence of fluctuations of thermal entanglement, $\Delta E(T)$, in the Heisenberg dimer.

Figure 3 shows the temperature dependence of entanglement and its fluctuations in the Heisenberg dimer. Entanglement $E(T)$ and its fluctuations $\Delta E(T)$ are equal to each other at the temperature $T_f$ determined from the relation

$$\frac{k_B T_f}{|J|} = 2/\ln\left[\frac{3(1 + C_f)}{1 - C_f}\right] = 0.57849.... \quad (33)$$

According to Fig. 3, entanglement fluctuations above temperature $T_f$ are larger than the entanglement itself. Near the entanglement disappearance point ($T = T_e$), the magnitude of the relative fluctuations diverges as $\delta E \sim 1/(1 - T/T_e)$. Figure 4 illustrates the behavior of $\delta E$ in the Heisenberg dimer.

At temperatures equal to or higher than $T_e$ (i.e., in the separable state), concurrence (31) and, thus, entanglement and its fluctuations vanish.

Thus, entanglement fluctuations should be taken into account for the entanglement entropy. Such fluctuations cannot be neglected near the points where the system transits from the entangled state to the separable one. The conceptions of entanglement fluctuations developed in this work can be useful in other fields where the concept of the entropy of the reduced density matrix is used.

We are grateful to the participants of the seminar "Quantum Computers" headed by Academician K.A. Valiev. We are especially grateful to A.S. Holevo for a number of fruitful remarks. This work is supported by the Russian Foundation for Basic Research (project no. 07-07-00048) and the Presidium of the Russian Academy of Sciences (program "Development of the Methods for Obtaining Chemical Substances and Creating New Materials").

## REFERENCES

1. W. K. Wootters, Quant. Inf. Comp. **1**, 27 (2001).
2. M. B. Plenio and S. Virmani, Quant. Inf. Comp. **7**, 1 (2007).
3. L. Amico, R. Fazio, A. Osterloh, and V. Vedral, Rev. Mod. Phys. **80**, 517 (2008).
4. R. Horodecki, P. Horodecki, M. Horodecki, and K. Horodecki, Rev. Mod. Phys. (in press); arXiv: quant-ph/0702225.
5. L. Bombelli, R. K. Koul, J. Lee, and R. D. Sorkin, Phys. Rev. D **34**, 373 (1986).
6. I. Klich and L. Levitov, Phys. Rev. Lett. **102**, 100502 (2009).
7. C. H. Bennett, G. Brassard, S. Popescu, et al., Phys. Rev. Lett. **76**, 722 (1996).
8. C. H. Bennett, H. J. Bernstein, S. Popescu, and B. Schumacher, Phys. Rev. A **53**, 2046 (1996).
9. L. D. Landau and E. M. Lifshitz, *Course of Theoretical Physics*, Vol. 5: *Statistical Physics* (Nauka, Moscow, 1995; Pergamon, Oxford, 1980).
10. H. Li and F. D. M. Haldane, Phys. Rev. Lett. **101**, 010504 (2008).
11. B. Nienhuis, M. Campostrini, and P. Calabrese, J. Stat. Mech.: Theory Exp., P02063 (2009).
12. C. H. Bennett, D. P. DiVincenzo, J. A. Smolin, and W. K. Wootters, Phys. Rev. A **54**, 3824 (1996).
13. S. Hill and W. K. Wootters, Phys. Rev. Lett. **78**, 5022 (1997).
14. W. K. Wootters, Phys. Rev. Lett. **80**, 2245 (1998).
15. K. Audenaert, F. Verstraete, and B. De Moor, Phys. Rev. A **64**, 052304 (2001).
16. A. A. Kokin, *Solid State Quantum Computers on Nuclear Spins* (Inst. Komp. Issled., Moscow, Izhevsk, 2004) [in Russian].
17. L. D. Landau and E. M. Lifshitz, *Course of Theoretical Physics*, Vol. 3: *Quantum Mechanics: Non-Relativistic Theory* (Nauka, Moscow, 1989, 4th ed.; Pergamon, Oxford, 1977, 3rd ed.).
18. S. M. Aldoshin, E. B. Feldman, and M. A. Yurishchev, Zh. Eksp. Teor. Fiz. **134**, 940 (2008) [JETP **107**, 804 (2008)].
19. M. A. Nielsen, "Quantum Information Theory," Dissertation (New Mexico, 1998); arXiv: quant-ph/0011036.
20. M. C. Arnesen, S. Bose, and V. Vedral, Phys. Rev. Lett. **87**, 017901 (2001).